# Thermal Conductivity of Cubic Silicon Carbide Single Crystals Heavily Doped by Nitrogen

Zifeng Huang[1], Yunfan Yang[2], Da Sheng[2], Hui Li[2, a)], Yuxiang Wang[1], Zixuan Sun[1], Ming Li[1], Runsheng Wang[1], Ru Huang[1], Zhe Cheng[1,3,a)]

[1] School of Integrated Circuits and Beijing Advanced Innovation Center for Integrated Circuits, Peking University, Beijing 100871, China

[2] Beijing National Center for Condensed Matter Physics and Institute of Physics, Chinese Academy of Sciences, 100190 Beijing, China

[3] Frontiers Science Center for Nano-optoelectronics, Peking University, Beijing 100871, China

a)Authors to whom correspondence should be addressed: zhe.cheng@pku.edu.cn; lihui2021@iphy.ac.cn

## Abstract

High-purity cubic silicon carbide (3C-SiC) possesses the second-highest thermal conductivity among large-scale crystals, surpassed only by diamond, making it crucial for practical applications of thermal management. Recent theoretical studies predict that heavy doping reduces the thermal conductivity of 3C-SiC via phonon-defect and phonon-electron scattering. However, experimental evidence has been limited. In this work, we report the thermal conductivity of heavily nitrogen-doped 3C-SiC single crystals, grown using the top-seeded solution growth (TSSG) method, measured via time-domain thermoreflectance (TDTR). Our results show that a significant reduction (up to 30%) in thermal conductivity is observed with nitrogen doping concentrations around $10^{20}$ cm$^{-3}$. A comparison with theoretical calculations indicates less intensive scatterings are observed in the measured thermal conductivity. We speculate that the electron-phonon scattering may have a smaller impact than previously anticipated or the distribution of defects are nonuniform which leads to less intensive scatterings. These findings shed light on understanding the doping effects on thermal transport in semiconductors and support further exploration of 3C-SiC for thermal management in electronics.

# 1. Introduction

Silicon carbide (SiC) is a promising wide-bandgap semiconductor material processing high break-down electric field, high thermal conductivity, and high electron mobility[1–3]. These properties position SiC as a potential leader in revolutionizing high-power devices and being used as thermal management materials[4–8]. Among the three commonly used polytypes of SiC—4H-SiC, 6H-SiC, and 3C-SiC—the cubic 3C-SiC stands out due to its simpler crystal structure, which is predicted to exhibit the highest thermal conductivity[9,10]. Additionally, 3C-SiC possesses the highest electron saturation velocity among these polytypes, making it highly attractive for power devices as electronic materials[2].

3C-SiC is also the only polytype that can be grown on a silicon substrate. As overheating becomes an increasingly critical issue in electronic devices, particularly in highly integrated silicon-based technologies like 3D integrated circuits, where the power density is high and heat dissipation paths are limited[11–14], the superior thermal performance and compatibility of 3C-SiC could provide an important solution as thermal management materials.

High-concentration nitrogen (N) doping is commonly used in the electronics industry to produce $n$-type SiC[15–17]. However, introducing impurities inevitably reduces thermal conductivity due to increased phonon scattering. Heavy nitrogen doping can significantly decrease thermal conductivity via defect-phonon and electron-phonon scattering mechanisms[10], which poses a challenge for thermal management in electronic devices. Despite the importance of this issue, detailed experimental data are still lacking.

Unlike 4H-SiC, the lower growth rate and quality of 3C SiC hinder further research on it [18,19]. Recently, the top-seeded solution growth (TSSG) method for 3C-SiC has been developed, enabling the growth of high-quality single crystals wafers while maintaining high N doping concentration up to $1.89 \times 10^{20}$ cm$^{-3}$ [20]. By utilizing this method, we can achieve both high crystal quality and substantial nitrogen doping, shedding light on the underlying principles of phonon scattering in heavily doped 3C-SiC. Understanding the mechanisms of thermal transport in this system, as well as the quantitative effects of doping, is crucial for optimizing growth conditions to balance electrical and thermal performance.

When N impurities are lightly doped, the impact of electron-phonon scattering on thermal conductivity is generally negligible. Previous experimental and theoretical studies support this view for 4H-SiC and 6H-SiC under light N doping conditions[10,21]. However, under heavily N doping situation, the theoretical work of Pang *et al*.[22] and Wang *et al*.[23] suggests that phonons can be significantly scattered by electrons excited from nitrogen donors, leading to a reduction in lattice thermal conductivity. Similar theories and experimental findings in silicon (Si) has been reported by Liao *et al*.[24,25] and Dongre *et al*.[26], who studied the coexistence of defect-phonon and electron-phonon scatterings in heavily doped Si. This phonon scattering mechanism is crucial for a fundamental understanding and broader application of 3C-SiC materials, though it has yet to be experimentally verified.

In this study, heavily N-doped 3C-SiC single crystals were grown on 4H-SiC substrates using the TSSG method[20]. Crystal type and quality was confirmed through Raman spectroscopy, while thermal conductivity was measured using time-domain thermoreflectance (TDTR)[27]. Additionally, secondary ion mass spectrometry (SIMS) was conducted to quantify the doping elements and their concentrations. A Callaway[28] model was used to extract the effect of Al impurity on thermal conductivity using Matthiessen's rule[29].

## 2. Materials and methods

2.1 Single crystal growth

Three heavily N-doped 3C-SiC samples were grown using the TSSG method. The Si-based solvents are loaded into a high-purity graphite crucible, both as the reaction container and the carbon source. An axial temperature gradient from the bottom to top is set to be~5–15 °C cm$^{-1}$ with the lowest temperature of ~1850°C at the top surface. The 4-inch 4H-SiC is used as the seed crystal with the (0001) surface as the growth front. The experimental details can be found in the Ref[20]. The samples tested in this study were grown under $N_2$ pressures of 25-30 kPa to achieve high concentration of N doping. The as-grown 3C-SiC ingots are then cut, grinded, and polished for the following characterizations. The wafers are cut into $10 \times 10$ mm$^2$ for the measurements. The roughness for the 3C-SiC wafers is ~1 nm.

2.2 Material characterizations

Raman spectroscopy was used to verify the structure of the samples. The measurements were performed using a Horiba LabRAM confocal Raman imaging system, equipped with a 50x objective lens. Spectra were acquired using 5% of the total laser power at a wavelength of 532 nm, with an acquisition time of 60 seconds over 5 accumulations. The peaks in the Raman spectra correspond to the energies of different phonon branches, confirming the crystal structure. Additionally, SIMS was used to determine the concentrations of N and other unintentional impurities. The SIMS analysis was conducted with equipment provided by Cameca Inc. More detailed structural analysis of the 3C-SiC crystals can be found in Ref [20].

2.3 Thermal characterizations

The thermal conductivity of the three samples was measured using TDTR, an optical thermal characterization method based on an ultrafast femtosecond laser. A Sapphire:Ti (Mai Tai) laser source, operating at 80 MHz with a wavelength of 785 nm and a pulse width shorter than 100 ps, generated a train of laser pulses. The laser beam was then split into a probe beam and a pump beam using a polarization beam splitter (PBS). The pump beam was modulated by an electro-optic modulator (EOM) at a specific modulation frequency, which was controlled by a signal from a function generator. The pump and probe beams were then recombined, and an objective lens focused them onto the sample surface with a defined spot size. A mechanical delay stage was used to adjust the arrival time of the pump beam relative to the probe beam. The pump beam induced periodic temperature variations on the sample's surface, and these thermal signals were detected by the probe beam via thermoreflectance. The reflected beam was captured by a photodiode and an RF lock-in amplifier. By fitting the experimental data to a theoretical heat transfer solution of the multilayer sample, the unknown thermal properties were obtained. In this study, a modulation frequency of 10.1 MHz and a spot radius of 12 μm were chosen to avoid any possible ballistic thermal transport phenomenon[9,30,31]. Further theoretical and experimental details on TDTR measurements can be found in Refs[27,32–35].

2.4 The Callaway model of thermal conductivity

According to the Callaway model of thermal conductivity, the thermal conductivity of a semiconductor, where heat conduction is primarily governed by phonons rather than electrons, can be expressed as [36]:

$$\kappa = \frac{1}{3} C_v v_a^2 \tau \tag{1}$$

where $\kappa$ is the thermal conductivity, $C_v$ is the volumetric heat capacity of the material, $v_a$ is the average phonon group velocity and $\tau$ is the phonon relaxation time. The product of phonon group velocity and phonon relaxation time is the phonon mean free path (MFP). The average phonon group velocity $v_a$ can be defined as:

$$\frac{1}{v_a^3} = \frac{1}{3}\left(\frac{1}{v_l^3} + \frac{2}{v_t^3}\right) \tag{2}$$

where $v_l$ is the longitudinal sound velocity and $v_t$ is the transverse sound velocity[37,38]. The effects of different scattering mechanisms—such as boundary, impurity, electron, and phonon scatterings—are encapsulated in the value of $\tau$. According to Matthiessen's rule, the total relaxation time $\tau$ is related to the individual relaxation time induced by each scattering sources can be expressed as[29,39,40]:

$$\frac{1}{\tau} = \frac{1}{\tau_{pd}} + \frac{1}{\tau_{pp}} + \frac{1}{\tau_{pe}} + \frac{1}{\tau_{pb}} \tag{3}$$

Where $\tau_{pd}$, $\tau_{pp}$, $\tau_{pe}$, and $\tau_{pb}$ represent the relaxation time of phonon-defect scattering, phonon-phonon scattering, phonon-electron scattering, and phonon-boundary scattering, respectively.

If multiple types of defects are present in the system, the relaxation time for phonon-defect scattering can be expressed as:

$$\frac{1}{\tau_{pd}} = \sum_i \frac{1}{\tau_{pd_i}} \quad (4)$$

The samples in this work are unintentionally doped with Al. To exclude the effect of the Al impurity, the corresponding scattering factor $\tau_{p-Al}^{-1}$ needs to be subtracted from the total scattering factors using equation (4). The corresponding thermal conductivity without Al impurity can be obtained.

## 3. Results and discussion

The Raman spectrum of the sample, shown in Fig. 1(a), features a distinctive Raman shift peak at 794 cm$^{-1}$, corresponding to the transverse optical (TO) phonon branch at the Γ point of 3C-SiC. The theoretical phonon dispersion relation for 3C-SiC, highlighting different branches, is depicted in Fig. 1(b). The Raman peak obtained in this study agrees well with reference values[41–43], confirming the polytype is 3C-SiC.

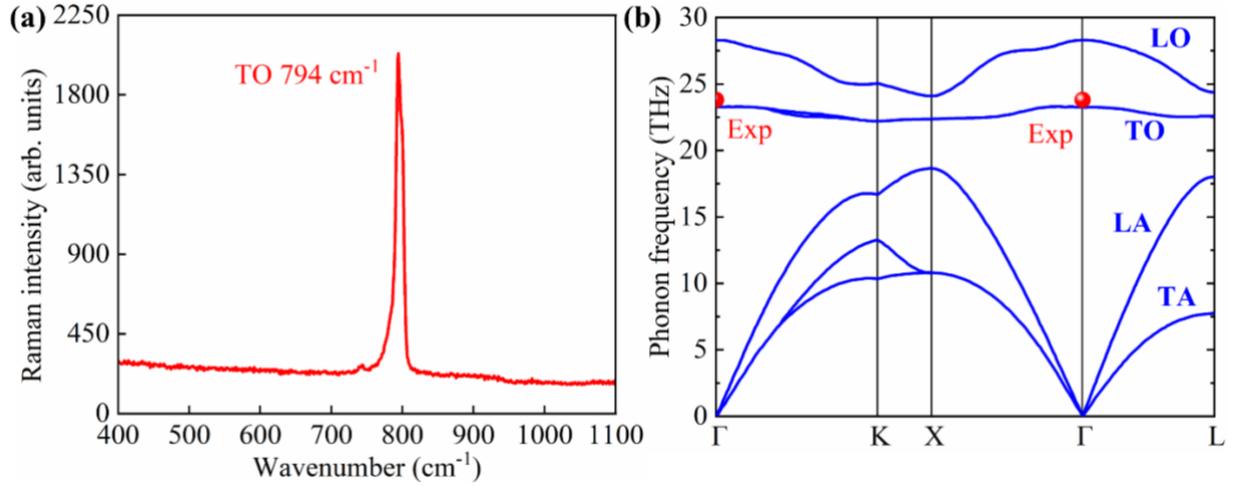

**Fig. 1.** Phonon energy as represented by the Raman spectra of the 3C-SiC single crystals. (a) The Raman spectra of the 3C-SiC sample from this study, with the peaks corresponding to different phonon branches. Only the peak of TO phonon branch can be observed, which is similar to Ref [20]. (b) The phonon dispersion relation of 3C-SiC, where the blue curve represents the theoretical results from Ref [23]. Different phonon branches are labeled as LO (longitudinal optical), TO (transverse optical), LA (longitudinal acoustic), and TA (transverse acoustic). The corresponding Raman peak of this sample is depicted in red circle.

Three samples, labeled as Sample A, Sample B, and Sample C, are studied in this work. Their doping conditions are determined by SIMS and the results are presented in Table 1. Nitrogen (N) is the dominant impurity in these samples with concentrations approximately two orders of magnitude higher than that of Al. However, according to pervious theoretical study[44], the low Al concentration reduces the thermal conductivity slightly. We separate the impact of N doping from the influence of Al by the Callaway mode described above.

Table 1. Doping concentrations of three 3C-SiC samples measured SIMS.

| Samples | A | B | C |
|---|---|---|---|
| N concentrations (atoms/cm$^3$) | 2.29×10$^{20}$ | 2.49×10$^{20}$ | 2.62×10$^{20}$ |
| Al concentrations (atoms/cm$^3$) | 5.65×10$^{18}$ | 2.12×10$^{18}$ | 8.27×10$^{18}$ |

Figure 2 shows the measured thermal conductivity of Samples A, B, and C with theoretical predictions based on different scattering theories. Figure 2(a) illustrates the TDTR ratio signals as a function of the delay time, and the best-fit curve of 3C-SiC $\kappa$, along with the calculated ratio curves with ±10% best fit 3C-SiC $\kappa$ which are depicted by two blue deashed lines. A significant difference between the red solid line and the blue dashed lines indicates the high sensitivity of the 3C-SiC $\kappa$ measurement.

To separate the impact of N impurities from unintentionally doped Al, a Callaway model is used with the Matthiessen's rule. The individual scattering rates due to Al impurities for the three doping concentrations are derived from the work of Katra *et al*. [44] by subtracting the undoped scattering rates from the doped scattering rates. By removing the Al scattering rates from the measured data, the thermal conductivity values without Al impurities are obtained, as indicated by the blue open triangles in Fig. 2(b). The measured results are shown as red circles, which are slightly lower than the derived values due to the presence of unintentionally doped Al impurity. Pervious experimental results of polycrystalline 3C-SiC ceramic from Ivanova *et al*. [45] was also included for comparison. The literature results deviate significantly from theoretical predictions due to phonon-boundary scattering induced by grain boundary and uncertainties in doping concentrations. The theoretical thermal conductivity which only considers phonon-defect scattering is plotted as green solid line[44] and the theoretical thermal conductivity which considers both phonon-defect and phonon-electron scatterings is plotted as black solid line[22] in Fig. 2(b). The results indicate that the influence of electron-phonon scattering due to N doping may be overestimated by Pang *et al*.[22], as it does not cause enough thermal conductivity reduction at doping concentrations on the order of $10^{20}$/cm$^3$. Another possible reason is that, at high doping levels, the distribution of N defects is not uniform at the atomic scale, resulting in less doped part and more heavily doped part, and lower the overall phonon scattering rate. In contrast, Pang *et al*. assumed a uniform distribution of N defects throughout the bulk crystal on atomic scale[22]. The overall impact of phonon-defect and phonon-electron scatterings in heavily doped semiconductors is more complicated than the assumption of the calculations.

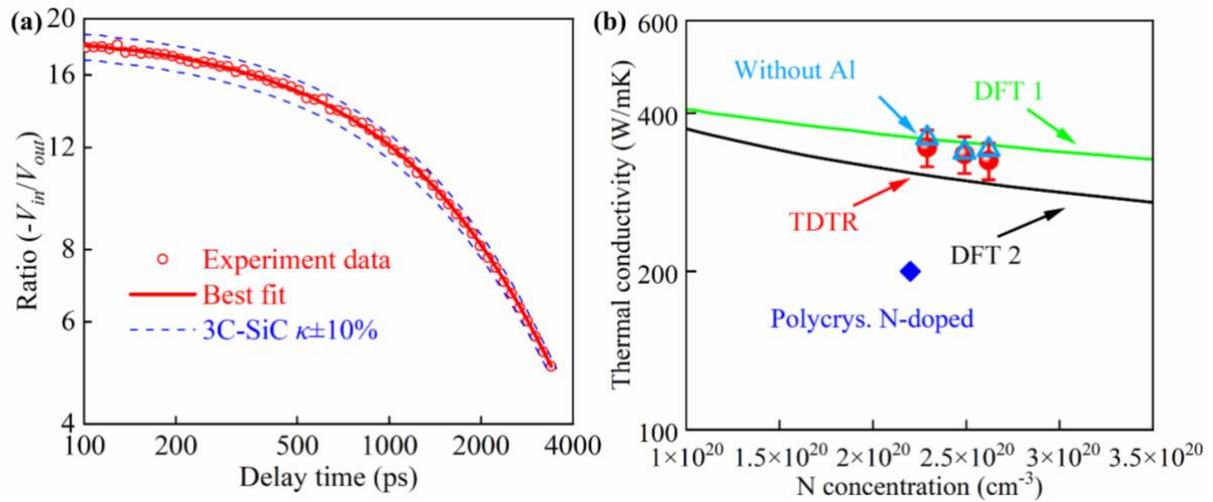

**Fig. 2.** The measured thermal conductivity of the three samples compared with previous experimental and theoretical results. (a) The typical ratio signal with the best fit curve, along with the curve with ±10% best fit 3C-SiC $\kappa$. (b) The experimental data from this study are shown as red circles, while the derived thermal conductivity values which exclude the impact of Al impurities are indicated by blue triangles. The measured thermal conductivity of polycrystalline N-doped 3C-SiC from Ivanova *et al.*[45] and the DFT calculated results from Katre *et al.*[44] (marked as DFT 1) and Pang *et al.*[22] (marked as DFT 2) are also included. The error bar of the measurement is ±8%.

To assess the overall uniformity of doping and crystal quality of the samples at macro-scale, 15 random spots were selected to measure thermal conductivity of 3C-SiC. The results, presented in Fig. 3, demonstrate good uniformity across the samples. The variation in thermal conductivity for each sample is less than 15%, with most data points fluctuating within approximately 10%, compared to the 8% systematic error bar of the TDTR measurement. This indicates that the thermal conductivity is uniformly distributed across the samples, supporting the conclusions discussed earlier and confirming the high and consistent quality of the 3C-SiC crystals. The nonuniformly doping mentioned above should be at the scale of nanometer, which is much smaller than the radius of the spot size (12 μm) in this study, so such variation cannot be detected by TDTR. The larger deviations in thermal boundary conductance (TBC) observed in Sample 3 are attributed to variations in the Al-SiC interfaces and are unrelated to the underlying 3C-SiC crystals.

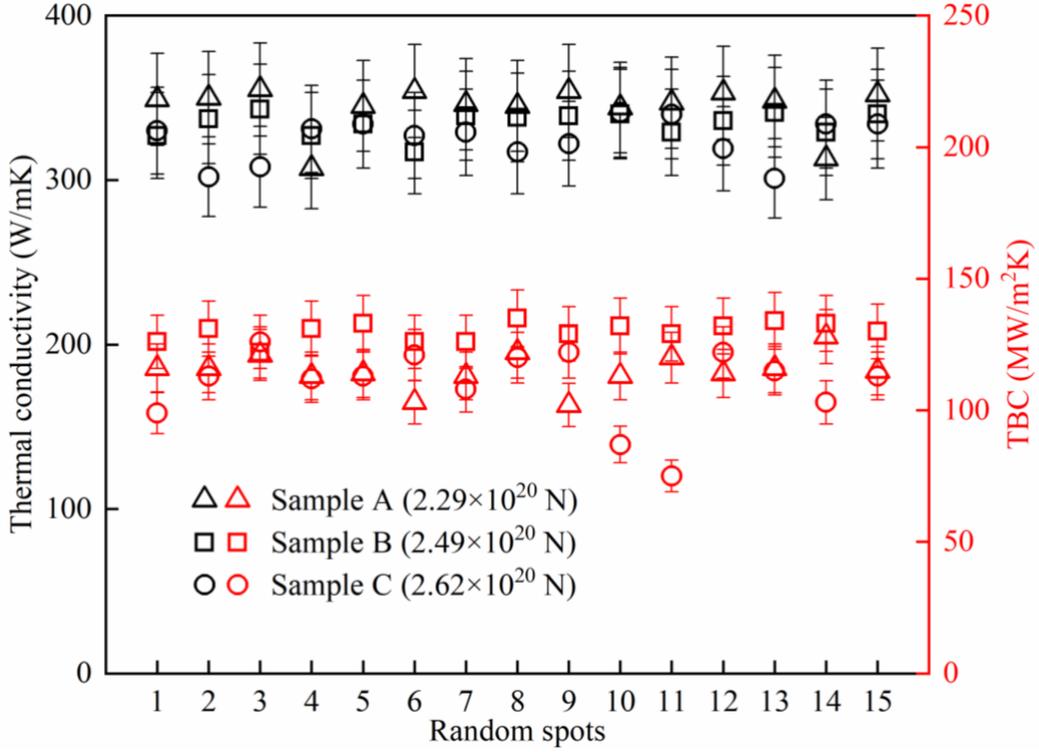

**Fig. 3.** The measured thermal conductivity and Al/3C-SiC thermal boundary conductance (TBC) at 15 random spots on each sample. Error bars indicate ±8% uncertainty. Black symbols represent thermal conductivity, while red symbols represent TBC.

To further investigate the thermal properties of heavily N-doped 3C-SiC, we performed TDTR measurements over a temperature range of 300-700K, as shown in Fig. 4, along with previous experimental and theoretical results. At high temperatures, phonon-phonon scattering dominates, while the defect-phonon scattering rate remains roughly constant, leading to a reduced impact of impurities on thermal conductivity[22]. Due to the similar doping concentrations of N impurities, the deviations among the measured thermal conductivity for the three samples are small. In logarithmic coordinates, the slope of each curve is approximately -1, indicating that at high temperatures, the $\kappa$ of all three samples is proportional to $T^{-1}$, suggesting that phonon-phonon Umklapp scattering is the dominant process in this temperature range[46].

The theoretical DFT prediction of the doping concentration of $1\times10^{20}$/cm$^3$ from Pang *et al.*'s work is depicted as black curve[22]. The measured thermal conductivity is lower than the predicted thermal conductivity near room temperature due to the higher concentrations of nitrogen. They converge at high temperatures because the predominant phonon scattering process is phonon-phonon Umklapp scattering. Consequently, for the same material with identical intrinsic anharmonic scattering properties, thermal conductivity converges to the same value at sufficiently high temperatures, regardless of doping conditions.

The experimental results for N-doped polycrystalline 3C-SiC ceramic from Ivanova *et al.*[45] are also included for comparison. The thermal conductivity measured in this study is much higher than Ivanova *et al.*'s results, as their samples are polycrystalline 3C-SiC ceramic with carbon mixed inside the samples, and the doping condition is inferred from carrier concentration, which is much lower than the actual doping concentration at room temperature. According to Pang *et al.*[22], achieving a carrier concentration of the order of $10^{20}$/cm$^3$ at room temperature requires an N doping concentration greater than $10^{21}$/cm$^3$, an order of magnitude higher than the concentrations in this study. Despite the poor crystal quality, the trend of thermal conductivity at high temperatures follows the same linearity but with a smaller absolute slope of about -0.6 above room temperature. The reduced thermal conductivity is primarily attributed to the extensive grain boundary scattering caused by the polycrystalline structure and the strong phonon-defect scattering due to high doping concentrations. Even at temperatures as high as 1000 K, the more intense phonon-phonon Umklapp scattering does not outweigh the effects of these two factors.

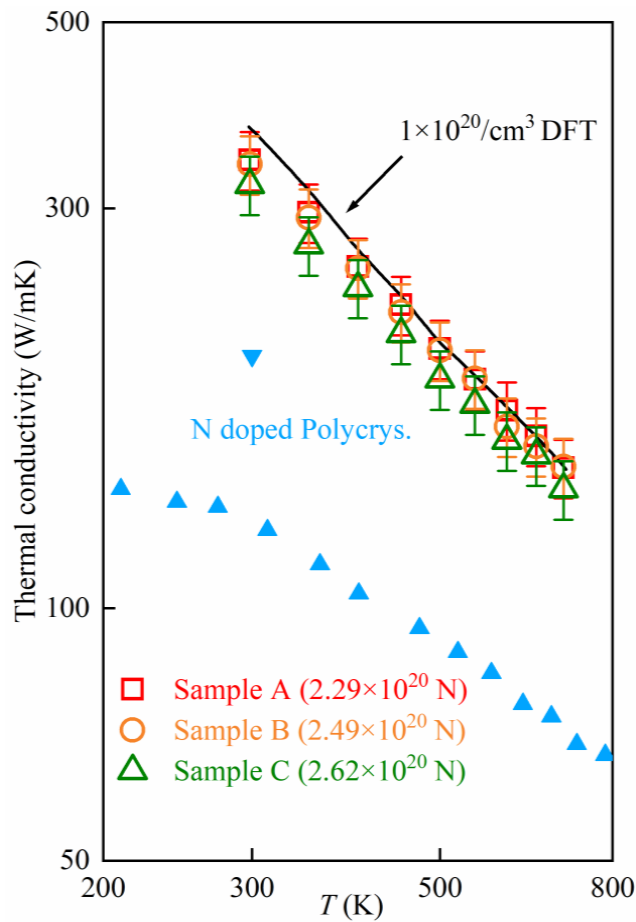

**Fig. 4**. The thermal conductivity of heavily N-doped 3C-SiC crystals at three different doping concentrations, measured over a temperature range of 300-700K, compared with previous experimental data at various doping concentrations (represented by discrete triangles). The experimental results from Ivanova *et al.*[45] are also included, where the upright triangles correspond to a carrier concentration of $2.2\times10^{20}/cm^3$, while the invers triangle corresponds to a carrier concentration of $8\times10^{20}/cm^3$.

## 4. Conclusions

In summary, this study measured the temperature dependent thermal conductivity of heavily N-doped 3C-SiC single crystals at various doping concentrations using TDTR. The effect of light Al impurities was removed by applying Matthiessen's rule. We also assessed the uniformity of doping and crystal quality by measuring thermal conductivity at random points, revealing good uniformity across the samples. The thermal properties at high temperatures were studied and compared with previous studies. The results are different from the calculations of Pang *et al*.[22]and aligns more closely with Katre *et al*.[44], where only defect-phonon scattering is considered. Our experimental results indicate less intensive phonon scatterings in the heavily doped semiconductors at doping concentrations around $10^{20}$/cm³. The electron-phonon scattering may be not strong as predicted or the possible nonuniform distribution of N defects leads to less intensive scatterings. These findings are important for 3C-SiC to be used as either active components in electronic devices or thermal management materials.

## Acknowledgements

Z.H., Y.W, and Z.C. thank Prof. Bo Sun for allowing them to use his TDTR system. The authors would like to thank Prof. Xiaolong Chen and Prof. Wu Li for their valuable comments and discussions. Z.H., Y.W, and Z.C. acknowledge the funding support from "The Fundamental Research Funds for the Central Universities, Peking University". Y.Y., D.S, and H.L. acknowledge financial support from the Beijing Municipal Science and Technology Project (Grant No. Z231100006023015).

## Competing interests

The authors declare no competing interest.

## Data availability

The data that support the findings of this study are available from the corresponding authors upon reasonable request.